\documentclass[twocolumn,aps,showpacs]{revtex4}
\usepackage[dvips]{graphicx}% Include figure files
\usepackage{dcolumn}% Align table columns on decimal point
\usepackage{bm}
\begin{document}
\draft  
\title{
Gradient-limited surfaces.
}

\author{Jaan Kalda}

\affiliation{
Institute of Cybernetics,
Tallinn Technical University,
Akadeemia tee 21,
12618 Tallinn,
Estonia
}

\begin{abstract} 
A simple scenario of 
the formation of geological landscapes is suggested and 
the respective lattice model is derived.
Numerical analysis shows that the arising non-Gaussian surfaces are characterized by the 
scale-dependent Hurst exponent, which varies from 0.7 to 1, in agreement with 
experimental data. 
\end{abstract} 
%05.40.-a   Fluctuation phenomena, random processes, noise, and Brownian motion
%91.10.Jf   Topography; geometric observations  
%68.35.Bs Structure of clean surfaces (reconstruction)
%64.60.Ak   Renormalization-group, fractal, and percolation studies of phase transitions 

\maketitle

Rough interfaces around us, such as Earth's surface  \cite{Mandelbrot2,Mandelbrot},  
surfaces of deposited films \cite{Meakin}, wetting fronts\cite{Buldyrev}, cloud perimeters \cite{Lovejoy},
fracture surfaces \cite{Bouchaud}, etc., are common objects, the properties and formation of which have been studied 
for several decades. In some cases, significant advances in theoretical understanding 
have been achieved. In particular, this applies to the surface growth processes, the
analysis of which has led to a wide variety of kinetic roughening models, cf.\ \cite{Meakin0,Kondev2}.
However, many processes leading to rough surfaces, eg.\ 
the formation of fractures and Earth's landscapes are less understood.

The formation of the Earth's surface is a complex process, affected by 
various phenomena,  such as seismic and tectonic activity, 
erosion, sedimentation, etc. These phenomena, in their turn, can be of diverse nature.
Thus, erosion can be  caused by meandering rivers, oceanic and atmospheric influence, by 
the motion of ice, avalanches, and so on. 
Furthermore, the physical properties of the ground 
vary in a very wide range. 
Incorporating all this diversity into a concise embraceable mathematical model is
a hopeless task. However, 
%$H \approx 0.6$ -- $0.9$, in a good agreement
%with experimental observations \cite{Mandelbrot2,Mandelbrot}.
the scale-invariant properties of the geologic landscapes 
appear to be surprisingly universal: in a reasonable approximation, they are typically
self-affine, with the Hurst exponent ranging between $H \approx 0.7$ and 
$0.9$ \cite{Mandelbrot2,Mandelbrot}.
Therefore, it is natural to expect that there is a simple, universal and robust
mechanism leading to such surfaces. 
In what follows we show that such a mechanism can be provided 
by the competition of erosion and tectonic activity: 
the model {\em of gradient-limited surfaces} incorporates these two effects in their simplest 
form and leads to realistic landscapes.

Most models of geological landscapes are based on the evolution of 
river networks \cite{Giacometti,Sinclair,Banavar,Chan}. The evolution of rivers plays undoubtedly an important role in the 
formation of landscapes, but is not able to increase the height of the mountains and will not
lead to self-affine surfaces. 
The only attempt of constructing a robust self-affine model of Earth's surface
has been made by Mandelbrot \cite{Mandelbrot2,Mandelbrot}.
He modelled roughening due to tectonic activity, and his method can be outlined as follows.
Inside a polygon (Earth's surface), a random point is coined. Through that 
point, a line of random direction is drawn. This ``fault line'' divides the polygon
into two parts, one of which is elevated (with respect to the other) by a unit height.
The procedure is repeated $N \to \infty$ times. The Brownian growth of  
height differences is eliminated by normalizing the surface height to $\sqrt N$.
This results in a self-affine surface with $H=0.5$. In order to address 
the discrepancy between the model and empirical values  $H \approx 0.7$ -- $0.9$,
the model has been generalized by replacing the Heaviside profile of the ``fault'' by 
a profile with singularity (so that the height change is given by $\Delta h = x^\alpha \mbox{sign} x$,
where $-0.5 < \alpha < 0.5$ and the $x$-axis is perpendicular to the ``fault line'').

The tectonic activity and formation of faults, as captured by the Mandelbrot's model, plays 
certainly an important role in the evolution of the Earth's surface.
Meanwhile, the singular shape of the fault profile is artificial, with no physical motivation.
Besides, there are no physical processes which would normalize 
the surface height to the number of faults. Instead of that, the
basic effect reducing the height differences is erosion.
As mentioned above, erosion itself is a very complex phenomenon which, in particular, has been addressed 
by the models of the evolution of river networks \cite{Giacometti,Sinclair,Banavar,Chan}. 
The excessively detailed erosion models, however, are not suited for revealing the most generic 
aspects of landscape roughening.
Therefore, we opt for the simplest possible approach and assume that effectively, erosion imposes
an upper limit to the modulus of the gradient of the surface height. More 
specifically, we assume that, as soon as a slope becomes steeper than a
threshold value, the excess of the height drop is spread over the neighboring regions.
Such a smoothing of too steep slopes can be accomplished, for instance, by avalanches.

To begin with, let us define the model of gradient-limited surfaces for a continuous medium.
A random point $P$ and direction $\bm {\tau}$ define a ``fault line'' inside a polygon of diameter $L \gg 1$.
This line divides the polygon into two parts, one of which (leftmost, with respect to the direction $\bm{\tau}$) is
elevated by a unit height. The height drop is spread over the nearest neighboring regions in such a way
%The actual elevation line follows the aim line as closely as possible in such a way that
%after the elevation, 
that the modulus of the local gradient remains everywhere below a threshold value, i.e.\ 
$|\bm{\nabla} \psi| < 1$. 
If the fault line goes through a region of a saturated slope, the avalanches can affect large areas. Then, the 
actual elevation (the change of surface slope) will take place far from the fault, at the edges of the saturated slope.
These edges will be referred to as the elevation lines.
%The elevation line is terminated as soon as it reaches the boundary of the polygon;
The procedure is repeated $N \to \infty$ times. Note that unlike the Mandelbrot's model, this model has
a lower cut-off scale (the ratio of the height drop at a single fault line and threshold gradient). 

The model can also be formulated on a lattice. A natural basis for the lattice formulation is given by the
six-vertex (restricted solid-on-solid) model \cite{Meakin,Rys}.
On the square lattice of the six vertex model, all the edges are marked with arrows so that each 
site has equal number of incoming and outgoing arrows, see Fig.~1. The arrows define an 
incompressible flow, the streamfunction of which is our surface. Thus, each arrow represents 
a unit jump in the surface height. The mean slope is defined by the mean density of counter-directed
arrows, which is limited by the step of the grid; hence, the slope steepness is constrained automatically.
\begin {figure}[!b]
\includegraphics{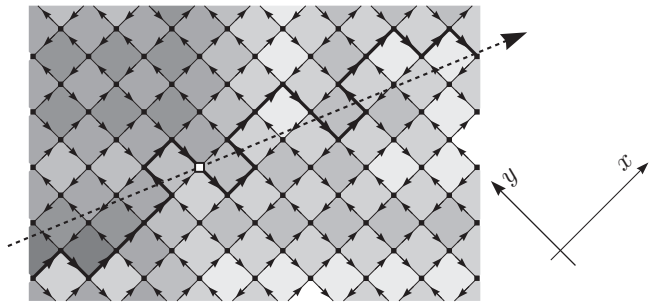}
\caption{The algorithm of finding the elevation line (depicted by bold line).
The aim line (dashed) is a random line through a randomly chosen (white)  vertex.
The elevation line is such a directed chain of arrows, which follows the aim line as closely as possible. 
Darker areas (squares) are lower.}
\end {figure}

Consider an oriented chain of arrows dividing the lattice into two parts (see Fig.~1). Swapping the 
direction of all the arrows of the chain is legitimate, because for all the affected sites, 
the number of incoming arrows is conserved. It corresponds to the lowering of one part of the 
surface with respect to the other part by two units. Therefore, {\em directed chains of arrows} can
play the role of {\em elevation lines}. Gradient-limited surfaces are obtained
as follows. A random site of the lattice $P$ and a random direction $\bm{\tau}$ are coined, they 
define an {\em aim line}, the site of the ``fault''.
That part of the surface, which is leftwards to the aim line, is to be elevated. 
The elevation is accomplished along such a directed chain of arrows, which follows as closely as possible
the aim line. Similarly to the continuous case, if the aim line goes through a region of uni-directional arrows 
(saturated slope), the elevation line is forced to go around those regions.
There are different technical options, how to minimize the distance between the elevation line and aim line.
In particular, the distance can be optimized locally and globally. However,
the scaling properties of the resulting surfaces are insensitive with respect to the particular choice.
For our main series of simulations, we used a local algorithm.
The elevation line was traced step-by-step, starting from the origin $P$. At each step, there are two 
possibilities to continue the line, because there are two outgoing arrows from each site (however, if the 
site has been already visited, one of the outgoing arrows is occupied, and there is only one possibility left).
That option is to be selected, which leads closer to the aim line. 
The (signed) departure from the aim line is easily tracked as $s = \Delta x \sin \alpha -\Delta y \cos \alpha$,
where $\Delta x$ and $\Delta y$ are displacements along $x$ and $y$ axes, and $\alpha$ --- the angle between the $x$-axis 
and the aim line. The elevation line is terminated as soon as it reaches the boundary of the polygon. 
The gradient-limited surfaces are obtained at the long-time limit, when the number of elevations exceeds
the relaxation time (which scales as the number of sites in the polygon, see below), 
and the initial shape of the surface becomes irrelevant.

\begin {figure}[!t]
\includegraphics{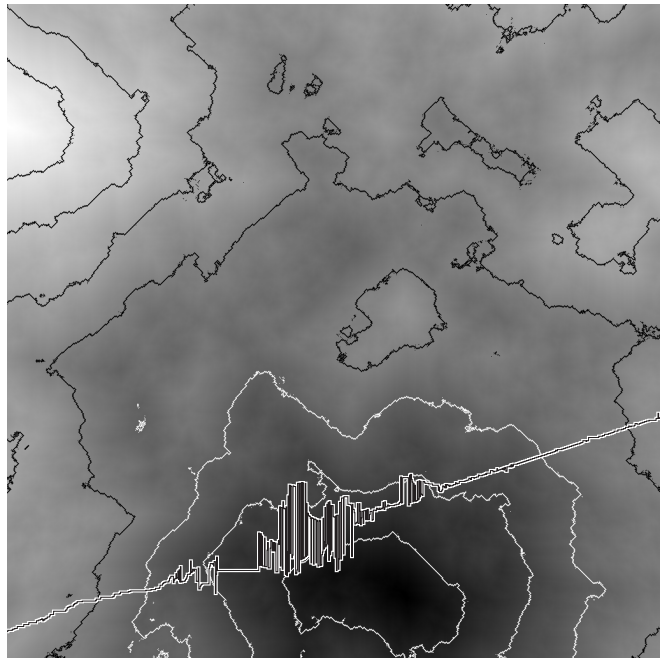}
\caption{A gradient-limited surface, polygon size $L_{\max}=2049$. Darker areas correspond to lower regions of the 
surface, black and white lines depict equi-distant  level lines. Black line surrounded by white is an elevation line.}
\end {figure}
A simulation result is presented in Fig.~2. 
Observe the shape of the elevation line: in the region of a saturated slope, it has to depart far from the straight aim line.
These regions are responsible for the long-range correlation of the surface 
height increments. Qualitatively, the saturated slopes are caused by accumulated excess of the
``faults'' of a certain direction. 
If there were no avalanches, that
excess would fluctuate as the square root of the number of ``faults'', tending to infinity.
Therefore, the presence of large saturated slopes should not be surprising. As it will be shown below, in one-dimensional (1D)
case,  the accumulation phenomenon gives rise to $H=1$ (i.e. typically, a saturated slope occupies the whole polygon). 
In 2D geometry, the accumulation effect is weaker and leads to a more interesting scaling behavior.

The one-dimensional (1D) version of the model is most conveniently formulated as a spin exchange problem, and
admits analytical solution. This analytic approach helps us to understand the features of the 2D-model.
Suppose there is a sequence of spins,  $\varphi_i = \pm 1$.
It is convenient to consider infinite periodic sequence, $\varphi_{i+N} = \varphi_i$, where $i \in Z$
and $N$ is the period. The spins $\varphi_i$ can be interpreted as the increments of a self-affine curve
$\psi_j = \sum_{i=0}^j\varphi_i$. A random point $k$ and a random spin increment $\nu=\pm 2$ 
are coined. The increment is to be added to the $k$-th spin, or, if it is not possible 
(resulting in $\varphi_k = \pm 3$), to the nearest suitable spin (i.e.\ to $\varphi_l=-\nu \pm 1$  with minimal 
value of $|l-k|$). If there is no suitable spin at all, the next pair of $k$ and $\nu$ are coined.
The procedure is repeated ad infinitum. Let us denote the relative number of positive spins by $\xi$.
Then, the height drop of the above defined self-affine curve $\psi_j$ at distance $N$ is $N|2\xi-1|$.
The quantity $\xi$ performs Brownian fluctuations, because  at each time step, it is randomly incremented by $\pm N^{-1}$.
At the limit $N\to \infty$, the probability density function $n(\xi,t)$
evolves according to the diffusion equation, 
$n_t = D n_{\xi\xi}$ with $D=(2N^2)^{-1}$ and no flux at the boundaries, $n_\xi(0,t) = n_\xi(1,t) = 0$.
The stationary solution $n(\xi,t) \equiv 1$ allows us to calculate the delta 
variance $\left<(\psi_{i+N}-\psi_i)^2\right> = N^2\int_0^1(2\xi-1)^2d\xi=N^2/3$,
which corresponds to $H=1$. The relaxation time of the spin exchange problem 
can be found as the diffusion time, $\tau \approx N^2$ (time is measured in the number of spin exchanges).
The relaxation time of the 2D gradient-limited surfaces can be estimated in the same way, because 
the height difference between left and right edges of the polygon performs also nearly-Brownian fluctuations.

\begin {figure}[!t]
\includegraphics{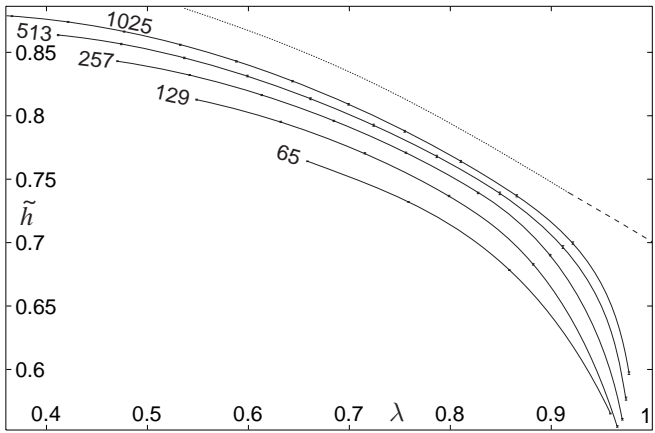}
\caption{Gradient-limited surfaces: numerical results. 
The numbers 1025, 513, 259, 127, and 65 indicate the edge length $L_{\max}$ of the polygon. The 
finite difference approximation of the differential roughness exponent
$\tilde h$ is plotted versus the logarithmic relative scale $\lambda$.
There is no strict self-affinity of the surface. However, at the limit $L_{\max} \to \infty$, there is 
an asymptotic dependance $\tilde h(\lambda,L_{\max}) \to \tilde h(\lambda)$ (dotted line, dashed part is extrapolation).}
\end{figure}
For two-dimensional geometry, the simulations indicate that 
the gradient-limited surfaces are not strictly speaking self-affine.
However, the data collapse of the simulation results can be achieved, when we introduce
the scale-dependent differential Hurst exponent, defined as
\begin{equation}
h(\lambda) = \frac 12 d \log \left< a_L^2 \right>/ d \log L, \;\;\;\;  \lambda = \log L /\log  L_{\max}.
\end{equation}
Here $a_L$ is the height of the surface at the distance $L$ from 
the center of the polygon of size $L_{\max}$. The angular braces denote averaging over different
realizations of the surface. In Fig.~3, the differential Hurst exponent 
is approximated by 
$\tilde h (\lambda,L_{\max})= \log (\left< a_i^2 \right>/\left< a_{i+1}^2 \right>) [\log (L_i /L_{i+1})]^{-1}$, where 
$i$ and $i+1$ are neighboring data-points and $\lambda = \log (L_i L_{i+1})/\log L_{\max}^2$. 
At the limit $L_{\max}\to \infty$, the curves 
converge to the asymptotic function $\tilde h (\lambda)$. 
Note that at the extreme right-hand-side of the plot, the convergence of the 
$\tilde h (\lambda,L_{\max})$-curves is not as good as elsewhere; this is explained by 
finite-size effects and by the fact that for $\lambda \approx 1$, the finite differences fail providing an
acceptable approximation for the derivative in Eq.~(1). 
The rapid fall-off of the curves at the limit $\lambda \to 1$ has the same origin, and therefore,
the values $\tilde h \alt 0.65$ are not reliable. For large scales (which are most interesting 
in the context of the Earth's surface) with $\lambda \agt 0.9$, a better approach is to extrapolate the asymptotic curve, see dashed line 
in Fig.~2. The conclusion $h \approx 0.7$--$0.9$ for $0.5 <\lambda <1$ is in a good agreement 
with the values recorded for geological landscapes, cf.\ \cite{Mandelbrot}.
Intriguingly, the roughness exponents of the fracture surfaces vary in the same range, cf \cite{Bouchaud,Sahimi,Maloy}.

\begin{figure}[!b]
\includegraphics{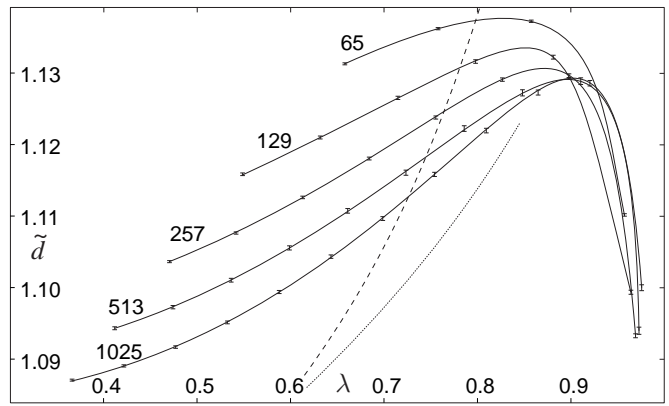}
\caption{Differential fractal dimension of the contour loops of gradient-limited surfaces: numerical results. 
The numbers 65--1025 indicate the polygon size $L_{\max}$. The finite difference approximation $\tilde d$ is 
plotted versus the scale $\lambda$. Dotted line depicts the asymptotic dependance $\tilde d(\lambda,L_{\max}) \to \tilde d(\lambda)$. 
Dashed line is calculated using the dependence $h(\lambda)$ (see Fig.~3), and the fractal dimension for Gaussian self-affine surfaces.}
\end{figure}
Finally, the scaling law of the overall (edge-to-edge) height drop of the gradient-limited surfaces is 
given by the integral Hurst exponent $H = \int_0^1 hd\lambda$: $\left< a(L_{\max})^2 \right> \propto L_{\max}^{2H}$; 
according to the simulations, $H = 0.91 \pm 0.01$.  This exponent equals to the area under the asymptotic curve in Fig.\ 3,
and this condition has been used to test the extrapolated curve in Fig.\ 3.

We have observed that the differential Hurst exponent is scale-dependent.
%; as it will be shown below, the fractal dimension of contour loops (``coastlines'') is also scale-dependent.
Such a generalized scale-invariance when critical exponents depend on scale, is not unique. 
For instance, similar behavior has been observed for certain forest fire models \cite{Chen}.
In our case, the Hurst exponent increasing towards small scales is caused by the 
presence of large areas of saturated slope. 
Indeed, consider a random  pair of points. If the distance $L$ between them is small, 
the points are likely to reside inside a single region of saturated slope. Hence,
their average height difference scales almost as $L$ implying $h \approx 1$.
On the other hand,  larger saturated slopes are more rare  than the smaller ones. 
Therefore, for a more distant pair of points, falling inside a single saturated slope is a rare event, and
the conclusion  $h \approx 1$ is no more valid.

For Gaussian self-affine surfaces, all the scaling exponents of statistical topography are functions of the 
Hurst exponent $H$. 
However, the gradient-limited surfaces are not Gaussian, as evidenced by the presence of large saturated slopes. 
Therefore, the differential Hurst exponent $h(\lambda)$ alone does not provide a complete 
description of the surface. First we consider the differential fractal dimension of the contour loops (``coastlines''),
$d= d \log \left< l_L \right>/ d \log L$. Here $l$ is the length of a contour loop, and $L$ its diameter.
The numerical results are given in Fig.~4.
For Gaussian surfaces, the fractal dimension of contour loops $D(H) \approx 1.5 - 0.5H$ \cite{Kondev2,JK}. 
Dotted line in Fig.~4 is the asymptotic ($L_{\max} \to \infty$) dependence $d(\lambda)$, and dashed line is 
the curve, calculated on the basis of the functions $h(\lambda)$ (from Fig.~3), and $D(H)$.
Evidently, $D(h(\lambda)) \ne d(\lambda)$.

\begin{figure}[!!t]
\includegraphics{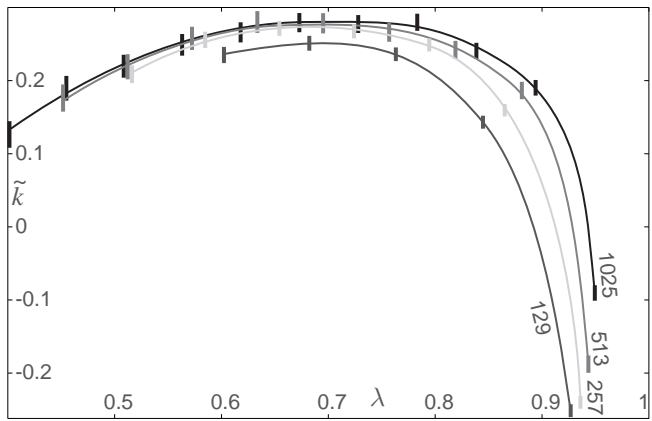}
\caption{The finite difference approximation $\tilde k$ of the differential scaling exponent characterizing the 
size-distribution of contour loops is plotted versus the scale $\lambda$. 
Positive values of $\tilde k$ indicate that there is an  anomalously small number of small 
contour loops (as compared with the Gaussian surfaces).}
\end{figure}
Even more pronounced mismatch between the Gaussian surfaces and the gradient-limited surfaces is observed
for size-distribution of the contour loops. Let $p(L)$ denote the probability that a randomly chosen point 
belongs to such a contour loop, the diameter of which is larger than $L$, but smaller than $2L$. 
Then, for Gaussian surfaces we would expect that $p(L) = L^k$, where $k=D(H)-(2-H)<0$ \cite{JK0,Kondev2}. For gradient-limited surfaces,
the curves $\tilde k(\lambda, L_{\max})$ converge again to the asymptotic dependence $\tilde k(\lambda)$.
This convergence with $\tilde k(\lambda)>0$ is clearly observed up to the scale $\lambda \approx 0.85$; above of that scale,
convergence is very slow due to finite-size effects.
The inequality $\tilde k>0$ means that, as compared with the Gaussian surfaces, there is significantly lesser number of small contour loops,
which is quite a natural observation. Indeed, the saturated slopes can be embraced by large contour loops, but leave almost 
no room for small ones.

In conclusion, the new model of gradient-limited surfaces leads to non-Gaussian surfaces which are 
characterized by the scale-dependent differential Hurst exponent. The latter varies from $h\approx 1$ for small scales, 
up to $h\approx 0.7$ for large  scales. This is in a reasonable agreement with the experimentally 
observed roughness of real geological landscapes. 
The relevance of the model to fracture surfaces deserves further analysis.

The support of ESF grant No 5036 is acknowledged. 
%The author is grateful to Prof.\ J.\  Engelbrecht for useful suggestions.

\end{document}